\documentclass[prl,twocolumn,showpacs,amsmath,amssymb,floatfix]{revtex4}

\usepackage{graphicx}

\begin{document}

\title{Decoherence Dynamics of Complex Photon States in a Superconducting Circuit}

\author{H. Wang}

\author{M. Hofheinz}

\author{M. Ansmann}

\author{R. C. Bialczak}

\author{E. Lucero}

\author{M. Neeley}

\author{A. D. O'Connell}

\author{D. Sank}

\author{M. Weides}

\author{J. Wenner}

\author{A. N. Cleland}
\email[]{anc@physics.ucsb.edu}

\author{John M. Martinis}
\email[]{martinis@physics.ucsb.edu}

\affiliation{Department of Physics, University of California, Santa Barbara, California, CA 93106}

\date{\today}

\begin{abstract}

Quantum states inevitably decay with time into a probabilistic mixture of classical states, due to their interaction with the environment and measurement instrumentation.
We present the first measurement of the decoherence dynamics of complex photon states in a condensed-matter system. By controllably preparing a number of distinct, quantum-superposed photon states in a superconducting microwave resonator, we show that the subsequent decay dynamics can be quantitatively described by taking into account only two distinct decay channels, energy relaxation and dephasing. Our ability to prepare specific initial quantum states allows us to measure the evolution of specific elements in the quantum density matrix, in a very detailed manner that can be compared with theory.

\end{abstract}


\maketitle

Quantum coherence provides the fundamental distinction between quantum and classical states, and is at the root of much of the non-intuitive behavior of quantum systems; quantum coherence also provides the central impetus for trying to build a quantum computer. Coherence is however extraordinarily delicate, decaying as a quantum system interacts with its environment, or vanishing as the result of even a simple measurement. A quantitative understanding of the decoherence process is critically important for predicting the behavior of quantum systems.  The harmonic oscillator is a particularly compelling system for such studies, as an oscillator includes an infinite number of energy levels, from which arbitrarily complex states may be created, yet theories for decoherence in this system are especially simple.
Harmonic oscillators are also generic, as they map into a large number of different physical systems: A highly useful example is the electromagnetic resonator, which is central to a number of approaches to building a quantum computer \cite{haro2006,schu2007,sill2007,maje2007,fink2008}, serving as a quantum memory element or a communication bus. The recent development of techniques for generating and measuring arbitrarily complex quantum states in a harmonic oscillator \cite{meek1996,varc2000,houc2007,guer2007,hofh2008,hofh2009} presents the intriguing possibility of exploring the detailed time evolution of these complex states, allowing testing of the theories of decay dynamics \cite{myat2000,dele2008,wang2008,brun2008,brun1996}.

Decoherence of complex harmonic-oscillator states has been studied in previous experiments \cite{myat2000,dele2008,brun1996}.  The effect of energy decay was measured using Schr\"{o}dinger cat states by observing loss of interference fringes \cite{myat2000,brun1996} or by directly filming the Wigner function \cite{dele2008}.  Dephasing was measured by injecting noise and then observing the loss of interference fringes for specially prepared states \cite{myat2000}.  Here we describe a series of experiments in which we prepare specific, complex, quantum-superposed photon states in an electromagnetic resonator, and then monitor the decay dynamics of its density matrix.  By selectively preparing certain types of states, we can highlight the decay of various elements in the density matrix, in particular off-diagonal ones describing phase coherence, for a precise comparison with theory.   For the first time, the time decay of various off-diagonal elements are explicitly measured and compared to theory, which show a unique and simple signature.  We also demonstrate that energy decay and dephasing can be measured simultaneously:  Surprisingly, dephasing rates can be extracted even when it is not dominant, indicating the precision of our measurement.   In characterizing a superconducting resonator circuit, our experiment measures an unexpected dephasing rate that is ~30 times slower than energy decay.  To perform these experiments, we have measured the decay of more than 13 unique off-diagonal elements, for the superposed photon Fock states $|0\rangle + |n\rangle$, $n=1$ to 8, for the states $|m\rangle + |3\rangle$, $m=1,2$, and for the Schr\"{o}dinger cat states.

Our experiment is performed using a half-wavelength coplanar waveguide resonator made from superconducting aluminum~\cite{hofh2008,wang2008,hofh2009}.  The cavity has a resonance frequency of 6.971 GHz, and is weakly coupled through a capacitor to an external microwave source.  Non-classical photon states are generated and measured via a second weak capacitive coupling to a superconducting phase qubit, whose state can be manipulated using a second microwave source, and measured quickly with single-shot fidelity near unity ($\sim90\,\%$)~\cite{luce2008}.  The strong non-linearity of the qubit allows complete control of the resonator photon state \cite{hofh2009}.  The qubit has an energy relaxation time $T_{1 \rm{q}} \sim 300\,\textrm{ns}$ and a phase coherence time $T_{2\rm{q}} \sim 120\,\textrm{ns}$, whereas the resonator has an energy relaxation time $T_1 \sim 2.4\,\mu\textrm{s}$, similar to previous devices \cite{neel20081,wang2008}.

A single measurement of the qubit state gives an outcome of $\left|g\right\rangle$ or $\left|e\right\rangle$; by repeating the state preparation and measurement, typically hundreds of times, we determine the probability $P_e$ for the qubit excited state.  Tomography of the resonator state involves measuring this probability for a range of resonator preparation parameters, followed by a photon swap with the qubit, as described below.

\begin{figure}
\begin{center}
\resizebox{0.45\textwidth}{!}{
\includegraphics[clip=True]{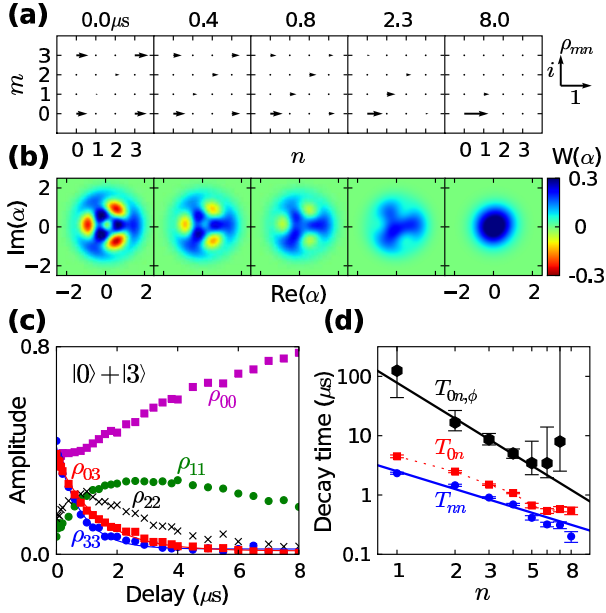}}
\end{center}
\caption{\label{fig.decay03} Decay of the $|0\rangle+|n\rangle$ states.
(a) Measured density matrices at five delay times for the decay of $|0\rangle+|3\rangle$.  The amplitude and phase of $\rho_{mn}$ is represented by the length and direction of an arrow in the complex plane (for scale, see legend on right).  The element $\rho_{03}$ ($= \rho_{30}^*$) is the only off-diagonal term that is non-zero during evolution, as expected.
(b) Reconstruction of the Wigner distribution W($\alpha$) from data in a.  The distribution decays to the ground state $|0\rangle$ for long times.
(c) Amplitudes of the density matrix elements versus time for the decay of $|0\rangle+|3\rangle$.  Solid lines are fits to the data for $\rho_{33}$ and $\rho_{03}$, from which decay times $T_{33}$ and $T_{03}$ are obtained.
(d) Plot of $T_{nn}$ (blue circles), $T_{0n}$ (red squares), and $T_{0n,\phi}$ (black hexagons) versus $n$ for $n=1$ to 8, obtained for a range of states $|0\rangle+|n\rangle$. Dephasing times $T_{0n,\phi}$ are calculated from Eq.~\ref{eq.0nT2}.  Blue line (slope = -1) and black line (slope = -2) are weighted least-square fits to obtain the resonator $T_1$ and $T_\phi$, respectively.}
\end{figure}

Non-classical resonator states are generated with qubit pulse sequences similar to those in previous experiments \cite{hofh2009}, which are based on the Law/Eberly protocol~\cite{law1996}. The resonator state is subsequently measured after a variable delay time $0\leq t \leq 15\,\mu\textrm{s}$ by first displacing the resonator with a coherent microwave pulse, which is characterized by a complex amplitude $\alpha$, where $|\alpha|^2$ corresponds to the average photon number.  The qubit is then brought into resonance with the resonator for a variable interaction time $\tau$, allowing photon exchange between the two, after which the qubit state is measured.  Because the resonator-qubit swapping frequency depends on photon number $n=0,1,2 ...$, a measurement of the excited qubit state $P_e$ versus $\tau$ is used to determine the photon occupation probabilities $P_n (\alpha)$ in the resonator.  In a prior publication \cite{hofh2009} we measured the full Wigner distribution $W(\alpha)$ from the parity $\sum_{n} (-1)^n P_n (\alpha)$ at thousands of points in the complex phase space $\alpha$.  Here we instead use a much more efficient sampling for $\alpha$, of only 60 points in total, arranged as two concentric circles with radii $|\alpha|$ = 1.10 and 1.45, from which we calculate the density matrix and reconstruct the Wigner distribution \cite{leib1996}.  The resulting readout fidelity is close to that obtained with full Wigner tomography, with the increased uncertainty ($\sim$2\%) mostly coming from the smaller sample size.

\begin{figure}
\begin{center}
\resizebox{0.45\textwidth}{!}{
\includegraphics[clip=True]{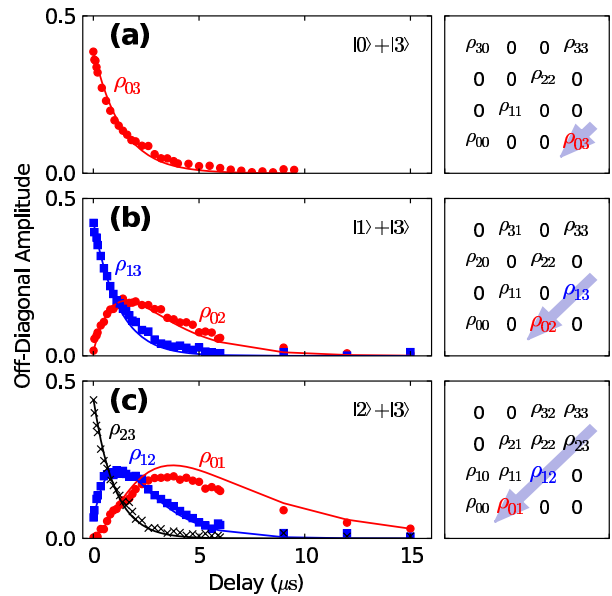}}
\end{center}
\caption{\label{fig.oddecay} Evolution of the off-diagonal elements for $\left|m\right\rangle+\left|3\right\rangle$
with (a) $m$ = 0, (b) $m$ = 1, and (c) $m$ = 2. Right panels illustrate the evolution of the off-diagonal elements.  Left panels display amplitude of matrix elements versus time.  Solid lines are simulations based on Eq.~\ref{eq.mass}, using $T_1 = 2.4 \,\mu\textrm{s}$ and $T_\phi = 70\,\mu\textrm{s}$. }
\end{figure}

The decay dynamics of a harmonic oscillator quantum state are described by a Markovian master equation that assumes uncorrelated energy relaxation and dephasing processes \cite{wall1985}.  The density matrix element $\rho_{mn}$ in the Fock state basis obeys
\begin{eqnarray}
\label{eq.mass}
\frac{d\rho_{mn}}{dt} & = & -\left[\frac{m+n}{2T_1}+\frac{(m-n)^2}{T_\phi}\right]\rho_{mn} \nonumber \\
& & + \frac{\sqrt{(m+1)(n+1)}}{T_1}\rho_{m+1,n+1} \ ,
\end{eqnarray}
where $T_1$ and $T_\phi$ are the resonator energy relaxation and dephasing times, respectively.  For $m=n$, Eq.~\ref{eq.mass} reduces to the master equation for photon number (Fock) state decay~\cite{wang2008}, with the $n$-photon Fock state lifetime given by $T_{nn}=T_1/n$.  For the off-diagonal elements $m\neq n$, dephasing causes $\rho_{mn}$ to decay at an additional rate proportional to the square of the distance from the diagonal.  In both cases, decay proceeds along the diagonal $\rho_{mn}$ $\rightarrow$ $\rho_{m-1,n-1}$ $\rightarrow$ $\rho_{m-2,n-2}$ $\rightarrow$ $\cdots$ (see Fig. \ref{fig.oddecay}).

\begin{figure}
\begin{center}
\resizebox{0.45\textwidth}{!}{
\includegraphics[clip=True]{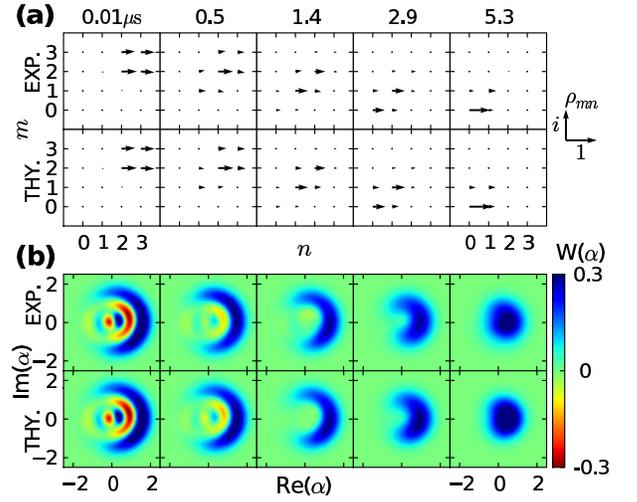}}
\end{center}
\caption{\label{fig.decay23} Snapshots of the time evolution of $\left|2\right\rangle+\left|3\right\rangle$.
(a) Measured and simulated density matrices at five delay times. The simulation starts with the measured density matrix at minimum delay and uses the decay parameters
$T_1 = 2.4\,\mu\textrm{s}$ and $T_\phi = 70\,\mu\textrm{s}$.  Note that only the off-diagonal elements $\rho_{23}$ $\rightarrow$ $\rho_{12}$ $\rightarrow$ $\rho_{01}$ are non-zero.
(b) Wigner distributions $W(\alpha)$ as a function of time, reconstructed from data in (a).}
\end{figure}

The decoherence for the Fock state superpositions $\left|0\right\rangle+\left|n\right\rangle$ is particularly simple, as only one independent off-diagonal element is non-zero.  This element, $\rho_{0n}$ ($= \rho_{n0}^*$), decays at the rate
\begin{equation}
\label{eq.0nT2}
\frac{1}{T_{0n}}=\frac{1}{2T_{nn}}+\frac{1}{T_{0n,\phi}} \ ,
\end{equation}
where $T_{0n,\phi}= T_\phi/n^2$.  By monitoring $\rho_{nn}$ and $\rho_{0n}$ for this type of state, we can directly determine the dephasing time $T_\phi$ and validate the prediction of Eq. \ref{eq.0nT2}.

In Fig. \ref{fig.decay03} we show the evolution of the state $\left|0\right\rangle+\left|3\right\rangle$ at five delay times; a full movie of 30 time steps is available online \cite{wang2009}.  Initially, the only non-zero terms in the density matrix are $\rho_{00}$, $\rho_{33}$, and $\rho_{03}$ ($= \rho_{30}^*$), as expected. As time evolves, the diagonal elements decay to $\rho_{00}$.  The off-diagonal element $\rho_{03}$ also decays with time, but without other off-diagonal states becoming occupied, as expected.
In Fig. \ref{fig.decay03}(c) we display the amplitudes of the $\rho_{33}$ and $\rho_{03}$ elements versus time, together with $\rho_{22}$, $\rho_{11}$ and $\rho_{00}$.
For clarity, statistical errors ($\sim$2\%) are not shown.  From the solid line fits we obtain the decay times $T_{33}$ and $T_{03}$, and using Eq. \ref{eq.0nT2} we extract the dephasing time $T_{03,\phi}$.

We have measured the evolution for the states $\left|0\right\rangle+\left|n\right\rangle$,  $n=1$ to 8; in Fig.~\ref{fig.decay03}(d), we show the measured decay times $T_{nn}$, $T_{0n}$, and $T_{0n,\phi}$.  We find $T_1 = 2.4\,\mu\textrm{s}$, with $T_{nn}$ scaling as $1/n$,  consistent with the $n$-photon Fock state lifetime, as observed previously~\cite{wang2008,brun2008}.
We also find that $T_{0n,\phi}$ has scaling consistent with $1/n^2$, confirming the dephasing mechanism. The resonator dephasing time $T_\phi$ is very long, approximately 70 $\mu$s, and is roughly consistent with off-resonant coupling to qubit dephasing.  Although $T_\phi \gg T_1$ and energy decay dominates decoherence, the precision of our measurement allows us to indentify a much smaller dephasing rate.

Decoherence theory can be further tested by creating states with non-zero elements closer to the diagonal, which then decay to other elements in the density matrix.  In Fig.~\ref{fig.decay23} we show the evolution for the state $\left|2\right\rangle+\left|3\right\rangle$, which initially has one non-zero off-diagonal element at $\rho_{23}$ ($= \rho_{32}^*$).  As the state evolves, this element decreases in amplitude, but generates non-zero amplitudes in the sequence of elements $\rho_{23}$ $\rightarrow$ $\rho_{12}$ $ \rightarrow$ $\rho_{01}$.

We plot the magnitude of these elements versus time in Fig.~\ref{fig.oddecay}, along with sequence data taken from the states $\left|1\right\rangle+\left|3\right\rangle$ and $\left|0\right\rangle+\left|3\right\rangle$.  The evolution is qualitatively similar to that found for the diagonal elements for Fock states \cite{wang2008}. We see good agreement between experiment and theory, indicating that the off-diagonal evolution is as predicted by the master equation. This is the first observation of the sequential movement of the off-diagonal coherence element in the density matrix due to energy relaxation.

\begin{figure}
\begin{center}
\resizebox{0.45\textwidth}{!}{
\includegraphics[clip=True]{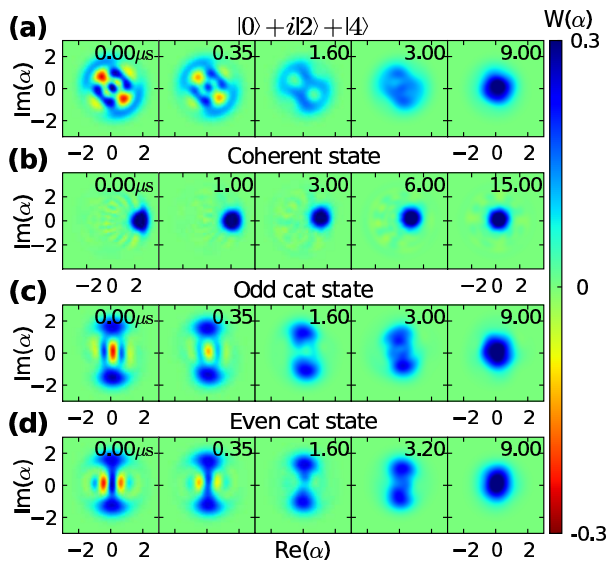}}
\end{center}
\caption{\label{fig.decayAll} Evolution of (a) the state $\left|0\right\rangle+i\left|2\right\rangle+\left|4\right\rangle$, (b) the coherent state  $\left|\alpha = \sqrt{5}\right\rangle$, (c) the odd Schr\"{o}dinger cat state $\left|\alpha = i\sqrt{2}\right\rangle - \left|\alpha = -i\sqrt{2}\right\rangle$, and (d) the even
Schr\"{o}dinger cat state $\left|\alpha = i\sqrt{2}\right\rangle + \left|\alpha = -i\sqrt{2}\right\rangle$.
Movies of the evolution for all states can be viewed online \cite{wang2009}. }
\end{figure}

We have also monitored the evolution of states involving more than a single pair of off-diagonal density matrix elements. Fig.~\ref{fig.decayAll}(a) shows the evolution of the state $\left|0\right\rangle+i\left|2\right\rangle+\left|4\right\rangle$, which initially has three unique non-zero off-diagonal elements. Fig.~\ref{fig.decayAll}(b) shows the decay of the coherent state $|\alpha =  \sqrt{5} \rangle$, created by driving the resonator with a classical pulse to an average photon number of
$\left|\alpha\right|^2=5$~\cite{hofh2008,wang2008}. During state evolution, the coherent peak is observed to move toward the origin at a rate consistent with $T_1$, while the Wigner distribution remains that of a coherent state, as expected from theory with $T_\phi \gg T_1$.

Figure~\ref{fig.decayAll} also shows the decay of the odd (c) and even (d) Schr\"{o}dinger cat states~\cite{dele2008}.  These states are created from a superposition of coherent states $\left|\alpha = i\sqrt{2}\right\rangle \pm \left|\alpha = -i\sqrt{2}\right\rangle$, where the even (odd) cat is for the plus (minus) sign, and consists of a superposition of even (odd) number photon states in the Fock-state basis.  In the experiment, the photon numbers are truncated above $n=6$.  As seen in the figure, the cat states feature a Wigner distribution with two peaks, corresponding to the coherent states displaced from the origin, but with quantum interference fringes between them. The even and odd cats have opposite fringe patterns, as expected. The interference fringes disappear at $\approx 1\,\mu$s, indicating the loss of phase coherence mostly due to energy relaxation.  The two coherent peaks begin to merge at $\approx 3\,\mu\textrm{s}$, indicating a complete energy decay of the system towards a vacuum state.

Movies of the decay dynamics for all the states mentioned here are available online \cite{wang2009}.  In all these states, movies constructed from the experimental data agree well with simulations based on the measured initial density matrix and the decay parameters $T_1 = 2.4 \,\mu\textrm{s}$ and $T_\phi = 70\,\mu\textrm{s}$.

In conclusion, we have measured the evolution of a number of non-classical photon states, prepared and measured in a microwave electromagnetic resonator using a superconducting phase qubit. The deterministically-generated states have non-zero off-diagonal elements in their density matrices, and we find that their time evolution is in excellent agreement with theoretical predictions based on the Markovian master equation, both in the sequence of elements and the individual decay rates.  While decay from energy loss is in good quantitative agreement with theory, we can only conclude that the decay due to dephasing is consistent with theory, due to the comparatively small effect of dephasing in this system 
($T_\phi \gg T_1$).  We also find that movies of decoherence for all the states we prepared and monitored are in excellent agreement with simulations.

\noindent \textbf{Acknowledgements.} Devices were made at the UCSB Nanofabrication Facility, a part of the NSF-funded National Nanotechnology Infrastructure Network. This work was supported by IARPA under grant W911NF-04-1-0204 and by the NSF under grant CCF-0507227.

\end{document}